\documentclass[useAMS,usenatbib]{mn2e}

\newcommand{\hii}{\hbox{H{\sc ii}}}
\newcommand{\etal}{\hbox{et~al.}}

\newcommand{\oiii}{\hbox{[O\,{\sc iii}]}}
\newcommand{\siii}{\hbox{[S\,{\sc iii}]}}

\title[]{Extreme CII emission in type 2 quasars at z$\sim$2.5: a
  signature of $\kappa$-distributed electron energies?}
\author[Humphrey et al.]{A. Humphrey$^{1}$, L. Binette$^{2}$\\
$^{1}$Centro de Astrof\'{i}sica da Universidade do Porto, Rua das
Estrelas, 4150-762, Porto, Portugal.  andrew.humphrey@astro.up.pt\\
$^{2}$Instituto de Astronom\'{i}a, Universidad Nacional Aut\'onoma de M\'exico, D.F., Mexico}

\begin{document}

\date{Accepted 2014 April 9.
      Received 2014 April 9;
      in original form 2014 January 24}

\pagerange{\pageref{firstpage}--\pageref{lastpage}}
\pubyear{2014}

\maketitle

\label{firstpage}

\begin{abstract}
We investigate the flux ratio between the 1335 \AA~ and 2326 \AA~
lines of singly ionized carbon in the extended narrow line regions of type 2
quasars at z$\sim$2.5. We find the observed CII $\lambda$1335 / CII]
$\lambda$2326 flux ratio, which is not sensitive to the C/H abundance
ratio, to be often several times higher than predicted by the
canonical AGN photoionization models that use solar metallicity and a
Maxwell-Boltzmann electron energy distribution. We study 
several potential solutions for this discrepancy: low gas metallicity,
shock ionization, continuum fluorescence, and
$\kappa$-distributed electron energies. Although we cannot
definitively distinguish between several of the proposed solutions, we
argue that a $\kappa$ distribution gives the more natural
explanation. We also provide a grid of AGN photoionization models
using $\kappa$-distributed electron energies. 

\end{abstract}


\section{Introduction}

One of the many advantages of looking to the high-redshift
Universe is the red-shifting of the rest-frame ultraviolet (UV)
emission into the optical observational regime, allowing us to access
the unique physics that can be probed using the UV emission or
absorption lines. 

In this paper we examine the ultraviolet CII $\lambda$1335 / CII]
$\lambda$2326 flux ratio as a diagnostic of the excitation
of extended narrow-line emitting gas associated with active galactic
nuclei. Our interest in this flux ratio
was initially motivated by the need for indicators of electron
temperature (T$_e$) in the rest-frame UV spectral region, to be used
with high-z galaxies for which the rest-frame optical temperature
diagnostics (i.e., [OIII] $\lambda$4363 / [OIII] $\lambda$5007) have
been red-shifted out of the optical observational regime. The large
difference in 
excitation energies of the two CII lines (9.3 vs 5.3 eV) makes their flux
ratio strongly sensitive to T$_e$, and hence the flux ratio potentially
offers a means to determine T$_e$. Furthermore, the use of lines from
a singly-ionized species ought to make this diagnostic readily
accessible for both high- and low-excitation objects, AGN and
star-forming objects, alike (cf. the [NeIV], [NeV] and OIII]
high-ionization temperature diagnostics discussed by Humphrey et
al. 2008: H08 hereinafter).  

In $\S$~\ref{cii_prob}~ we discuss the observational data used in this
paper: the CII $\lambda$1335 / CII] $\lambda$2326 flux ratio in type
2, radio-loud quasars at z$\sim$2.5. These data are compared against
different excitation models in $\S$~\ref{models}, and in
$\S$~\ref{ngc1068}~ we discuss the case of the narrow line
region of the nearby Seyfert 2 galaxy NGC 1068, which has been
extensively observed at UV wavelengths. In $\S$~\ref{disc}~ we
summarize and discuss our findings, and in appendix A1, we give a
subset of the exploratory grid of models, using $\kappa$-distributed
electron energies, that were computed during the preparation of this
paper; the full grid has been made available online. 

\begin{figure}
\includegraphics{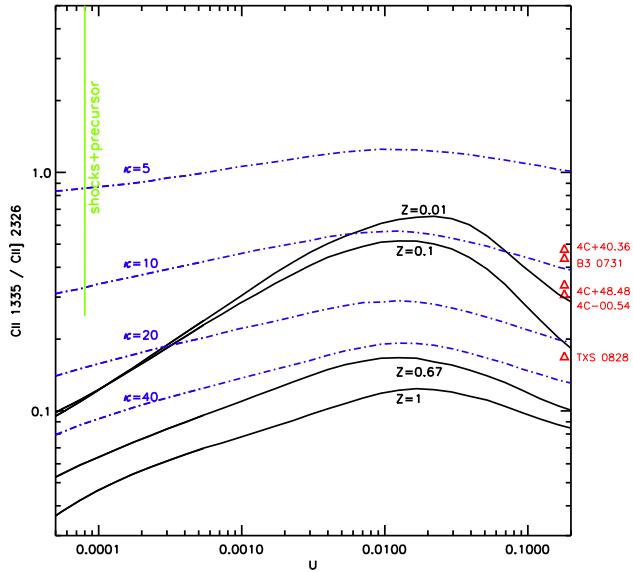}
\vspace{3.5in}
\caption{Results of AGN photoionization model calculations, showing
  the CII $\lambda$1335 / CII] $\lambda$2326 flux ratio vs. ionization
parameter U. The loci of the photoionization models that use a
Maxwell-Boltzmann distribution of electron energies are shown by 
solid black lines, and are plotted for gas metallicities of 1.0, 0.67,
0.1 and 0.01 times the solar value. Loci of photoionization models
using solar gas metallicity together with $\kappa$-distributed
electron energies are shown by dot-dashed 
blue lines, and are plotted for $\kappa$ = 40, 20, 10 and 5. In the
interest of clarity, the sequences with $\kappa$ = 2.5 and 80 are not
shown in this figure. The
range of shock model predictions of Allen et al. (2008), plotted at an
arbitrary value of U, are shown by the green vertical line. The red
triangles show the CII $\lambda$1335 / CII] $\lambda$2326 flux ratios
measured from the type 2 quasars, also plotted at arbitrary values of U.}
\label{fig_models1}
\end{figure}

\section{Data from the literature}
\label{cii_prob}

For this investigation we use line fluxes from the spectropolarimetric
study of 9 radio-loud type 2 quasars (also known as radio galaxies),
at z$\sim$2.5 published by Vernet et al. (2001:  
V01). Of those 9 galaxies, V01 detected 5 of them in both CII
$\lambda$1335 and CII] $\lambda$2326. The fluxes and line ratios of
these two lines are listed in Table ~\ref{cii_obs}. Where both lines
have been measured, the CII $\lambda$1335 / CII] $\lambda$2326 flux
ratio occupies the range 0.18$\pm$0.02 - 0.48$\pm$0.07.

\section{Comparison against models}
\label{models}

\subsection{Canonical AGN photoionization models}
\label{reference_grid}
As has been shown in several previous studies, the UV-optical emission
line ratios measured from the narrow line emitting nebulae associated
with high redshift type 2 quasars can, in general, be well reproduced
using photoionization models that use roughly solar gas
metallicity\footnote{For solar chemical abundances, we adopt the
  recent determination by Asplund et al. (2006).} ,
and an ionizing spectrum defined by a power-law of $S_v \propto
v^{-1.5}$ (e.g., Villar-Mart\'{i}n et al. 1997, V01, H08). These
models traditionally assume a Maxwell-Boltzmann equilibrium
distribution (MBED hereinafter) of electron energies in the ionized
plasma. 

As a starting point, we compute a sequence of models
with the above parameters using the photoionization code Mappings 1e
(Binette, Dopita \& Tuohy 1985; Ferruit et al. 1997; Binette et
al. 2012). We adopted a plane-parallel, single-slab geometry with a
constant hydrogen density of 100 cm$^{-3}$. Each model computation
was terminated when the hydrogen ionization fraction fell below 1 per
cent of its maximum. The ionization 
parameter U was varied along the model sequence from
1$\times$10$^{-5}$ to 0.20.  In this
solar-metallicity model
sequence, the maximum value for the CII $\lambda$1335 / CII]
$\lambda$2326 flux is 0.11, occurring at U$\sim$0.01
(Fig. ~\ref{fig_models1}).  This is a factor of 3 lower than the
observed mean value (Table ~\ref{cii_obs}), showing that the canonical
AGN photoionization models for the extended narrow line emitting gas
fail to adequately reproduce the observed CII $\lambda$1335 / CII]
$\lambda$2326 ratios. 

We note that the models also under predict the ratio of CII
$\lambda$1335 relative to several other lines, including
Ly$\alpha$, by a similar factor. Furthermore, the low ionization
resonant lines SiII $\lambda$1265 and SiII $\lambda$1309 are similarly
under predicted relative to CII] $\lambda$2326 and Ly$\alpha$. 
However, these particular line ratios are additionally subject to
relative chemical abundance effects, and hence do not provide as
strong a diagnostic power as the CII $\lambda$1335 / CII]
$\lambda$2326 ratio.  For this reason, for the remainder of this paper
we will focus on the CII ratio.  

While the CII ratio is also sensitive to reddening by dust,
any de-reddening would result in an even larger discrepancy between
the model and observed CII $\lambda$1335 / CII] $\lambda$2326 ratios. 
We also point out that the CII $\lambda$1335 / CII] $\lambda$2326 flux
ratio is essentially insensitive to the C/H abundance ratio.

\subsection{Sub-solar gas metallicity}
\label{subsolar}
Next, we investigate whether sub-solar metallicity ($Z$ hereinafter)
in the photoionized gas can help to explain the CII $\lambda$1335 /
CII] $\lambda$2326 
ratios.  We have repeated the above photoionization calculations, to
obtain model sequences at $Z/Z_{\odot}$ of 0.5, 0.25 and 0.1. All
metals were scaled linearly from solar abundances except for nitrogen,
which we scaled quadratically (N/O $\propto$ O/H; see e.g. Henry,
Edmunds \& K\"oppen 2000; Vernet et al. 2001) from its solar
abundance down to $Z/Z_{\odot}$=0.25, below which we scaled linearly
(N/H $\propto$ O/H). We find that 
decreasing the gas metallicity yields higher CII 
$\lambda$1335 / CII] $\lambda$2326 ratios, due to the increase in
electron temperature as cooling via emission lines becomes less
efficient at lower $Z$. Consequently, using sub-solar metallicity
allows the observed flux ratios to be reproduced
(Fig. ~\ref{fig_models1}). 

While the degeneracy between $Z$ and U makes it difficult to obtain a
precise value of $Z$ for each object, we can nonetheless use the
maximum value of CII $\lambda$1335 / CII] $\lambda$2326 at a given $Z$ 
to obtain an upper limit on metallicity $Z$. The observed values of the line ratio
require $Z/Z_{\odot}\la$0.25, the exception being TXS 0828+193, for
which we estimate $Z/Z_{\odot}\la$0.8. Taken at face value, such low
gas metallicities are at odds with the 
values $Z/Z_{\odot}\sim$1 for several of these type 2
quasars, as estimated in previous studies using the nitrogen lines
such as NV $\lambda$1240 or [NII] $\lambda$6583 
(see V01 and H08).

\begin{table}
\centering
\caption{The observational data used in this paper, from Vernet et
  al. (2001).  Fluxes are given in units of 10$^{-17}$ erg s$^{-1}$ cm$^{-2}$.} 
\begin{tabular}{lllll}
\hline
Galaxy & z & CII $\lambda$1335 & CII] $\lambda$2326 & $\lambda$1335 / $\lambda$2326 \\
\hline
B3 0731+438 & 2.43 & 2.2$\pm$0.2 & 5.0$\pm$0.5  & 0.44$\pm$0.06 \\
TXS 0828+193 & 2.57 & 3.3$\pm$0.3 & 19.2$\pm$1.9 & 0.17$\pm$0.02 \\
4C-00.54 & 2.36 & 1.1$\pm$0.1 & 3.6$\pm$0.4 & 0.31$\pm$0.04 \\ 
4C+40.36 & 2.27 & 20.9$\pm$2.1 & 43.3$\pm$4.3 & 0.48$\pm$0.07 \\
4C+48.48 & 2.34 & 3.9$\pm$0.4 & 11.6$\pm$1.2 & 0.34$\pm$0.05 \\
\hline
mean & & & & 0.35 \\
median & & & & 0.34 \\
\hline
\end{tabular}
\label{cii_obs}
\end{table}

\subsection{$\kappa$-distributed electron energies}
\label{kappa}
Recently, Nicholls et\,al. (2012, 2013) explored the possibility that electrons in \hii\ regions and PNe depart from a standard Maxwell-Boltzmann equilibrium energy
distribution. To represent non-Maxwell-Boltzmann electron energy
distributions, these authors adopted the parametrization given by a
$\kappa$--distribution, which is a generalized Lorentzian distribution
initially introduced by Vasyliunas (1968). It is widely used in the
studies of solar system plasmas (Livadiotis \& McComas 2011;
Livadiotis \etal\ 2011) where evidence of suprathermal electron
energies abounds. Nicholls \etal\ found that $\kappa$--distributed
electron energies resolved many of the difficulties encountered when
attempting to reproduce the temperatures observed in nebulae. 

The $\kappa$--distribution is a function of temperature and
$\kappa$. When compared to the standard MBED, it results in an excess
of electrons (or suprathermal tail) at high energies. The smaller
$\kappa$, the larger the excess becomes. This has the effect of
substantially enhancing the excitation rates of lines with a high
excitation energy as well as reducing the recombination rates (see
detailed study of Nicholls \etal\ 2013).  In the limit as $\kappa
\rightarrow \infty$, the distribution becomes a MBED. Note that the
electron mean energy is the same for a $\kappa$ distribution as for a
MBED ($3/2\,kT_e$). Binette \etal\ (2012) found that a
$\kappa$--distribution better reproduces the excess in \oiii\
temperatures over that of the 
\siii\ temperatures, which is found in \hii\ galaxies, giant
extragalactic \hii\ regions, Galactic \hii\ regions, and \hii\ regions
from the Magellanic Clouds. Nicholls \etal\ proposed various
mechanisms that would result in a suprathermal distribution, such as
(1) magnetic reconnection followed by the migration of high-energy
electrons along field lines, and by the development of inertial
Alfv\'en waves, (2) the ejection of high-energy electrons from the
photoionization process itself (when the ultraviolet source is very
hard, as in PNe or AGN), (3) or from photoionization of dust (Dopita
\& Sutherland 2000), and (4) by X-ray ionization, resulting in highly
energetic ($\sim 1\,$keV) inner-shell (Auger process) electrons (e.g.,
Aldrovandi \& Gruenwald 1985).

Mappings 1e allows the usage of a $\kappa$ distribution of electron
energies in place of the standard Maxwell-Boltzmann distribution
(see Binette et al. 2012).  We have computed a grid of models using
identical parameters to those used for our canonical model grid
described in $\S$~\ref{reference_grid}, but with $\kappa$-distributed
electron energy distributions corresponding to
$\kappa$ = 2.5, 5, 10, 20, 40, 80. 

As described by Nicholls et al. (2012, 2013) and Binette et
al. (2012), one of the effects of using a $\kappa$ distribution is the
general relative enhancement of collisionally excited lines with higher
excitation energies. As shown in Fig. ~\ref{fig_models1}, this is
indeed the case for the CII $\lambda$1335 / CII] $\lambda$2326 flux
ratio. In order to reproduce the observed flux ratios, we require
10$\la\kappa<\la$40. This is broadly consistent with
the range 10$<\kappa<$20 deduced by Nicholls et al. (2012) for
stellar-photoionized HII regions and planetary nebulae. 

A subset of our $\kappa$ distribution model grid is shown in Appendix
A1, with the complete grid made available online. 

\subsection{Shock models}
In addition to photoionization by the radiation field of the active
nucleus, fast shocks driven by the radio jets have also been proposed to
contribute to the ionization of the extended narrow line emitting gas
in radio-loud active galaxies of the kind considered herein
(e.g. Sutherland et al. 1993; Best, Longair \& R\"ottgering 2000; De
Breuck et al. 2000; Bicknell et al. 2000; 
Inksip et al. 2002; H08). 

In Fig. ~\ref{fig_models1} we also show the range of 
the CII $\lambda$1335 / CII] $\lambda$2326 ratio produced by the
shock models of Allen et al. (2008), for shocks velocities ranging
from 100 to 1000 km s$^{-1}$, with $B/n^{1/2}$=3.23 $\mu$G cm$^{3/2}$
and solar chemical abundances. For the line ratio under consideration,
the models show no significant difference between the shock only and
shock with precursor models, and thus we show only the latter. In the
interest of clarity, the shock models have been plotted at an
arbitrary value of U.  

From Fig. ~\ref{fig_models1} we can see that the shock models predict
significantly higher values for the CII $\lambda$1335 / CII]
$\lambda$2326 flux ratio than are predicted by our canonical
photoionization models. With the exception of TXS 0828+193, the shock
models are able to reproduce the measured CII ratios. 

\subsection{Continuum Fluorescence through CII $\lambda$1335}
\label{res}
The CII $\lambda$1335 / CII] $\lambda$2326 flux ratio can also be
affected by resonance scattering. This is because the observed flux of
a resonant line (i.e. CII $\lambda$1335 ) can be diminished or
enhanced relative to non-resonant lines (i.e., CII] $\lambda$2326),
depending on the gaseous geometry of a system (e.g. Villar-Mart\'{i}n,
Binette \& Fosbury 1996).

Indeed, recent detections of linearly polarized Ly$\alpha$ emission
from large-scale gaseous nebulae at high redshift have shown
that Ly$\alpha$ or continuum fluorescence, via illumination of neutral
hydrogen by a 
central source, is partly responsible for some of the Ly$\alpha$
emission for at least a fraction of giant Ly$\alpha$ nebulae (Hayes,
Scarlata \& 
Siana 2011; Humphrey et al. 2013; see also Prescott et al. 2011). This
process can also be expected to result in the continuum fluorescence
through other 
resonant lines, provided that the relevant species are present and
that the illuminating radiation field contains photons at the
wavelength of the relevant line. 

To accurately calculate the luminosity of CII $\lambda$1335 photons arising
by continuum fluorescence, detailed knowledge would be needed of
the column density and covering factor of singly ionized carbon, in
addition to the continuum luminosity of the quasar at ~$\sim$1335 
\AA, and would also require careful consideration of the radiative
transfer of the CII $\lambda$1335 through the gaseous environment of
the galaxies. Nevertheless, we can provide a ball-park estimate using some
simplifying assumptions. 

We consider a partially obscured, luminous quasar, which
illuminates gas in its host galaxy via a pair of diametrically opposed
ionization cones with a total opening angle of $\sim$2 Sr. Assuming
the hidden quasar is near the top of the luminosity 
function, we adopt a specific luminosity of $\sim$5$\times$10$^{43}$ erg
s$^{-1}$ \AA$^{-1}$ near the wavelength of CII $\lambda$1335, which
corresponds to a flux density of $\sim$3$\times$10$^{-16}$ erg
s$^{-1}$ cm$^{-2}$ \AA$^{-1}$ at z$=$2. We adopt a gaseous geometry
such that one third of sight-lines to the quasar have a radially
integrated column density of singly ionized carbon of $\sim$10$^{14}$
cm$^{-2}$, with no gas along all other sight lines. 

Evaluating the total
flux of the continuum photons that scattered through the CII
$\lambda$1335 line, we then obtain
$\sim$10$^{-17}$ erg s$^{-1}$ cm$^{-2}$ -- the same order of magnitude
as the observed CII $\lambda$1335 fluxes in the type 2 quasars (Table
~\ref{cii_obs}). Clearly, our estimate is subject to large
uncertainties, but this exercise does show that continuum fluorescence 
via AGN continuum illumination is able to make a potentially significant
enhancement to the luminosity of the CII $\lambda$1335 lines in the
host galaxies of powerful AGN. 

In addition to CII $\lambda$1335 (and Ly$\alpha$), this effect could
potentially enhance other resonant lines. Lines from astrophysically
abundant elements, which have a low collisional excitation probability
together with a high oscillator strength (e.g. OI $\lambda$1302, CI
$\lambda$1277 or CI $\lambda$1560), are expected to show the
highest relative flux enhancements.  

\section{The NLR of NGC 1068}
\label{ngc1068}
In addition to luminous active galaxies at high redshifts,
relatively nearby Seyfert galaxies offer an alternative
window by which the ultraviolet spectrum of extended, narrow line
emitting gas may be viewed. Among these, the nearby Seyfert 2 galaxy
NGC 1068 is one of the most extensively studied, and its narrow line
region has been the subject of several of ultraviolet
spectroscopic observations using space-based 
telescopes, which have revealed a spectrum rich in lines from a wide
range of metallic species, as well as recombination lines from HI and
HeII (e.g., Snijders et al. 1986; Kriss et al. 1992; Kraemer et
al. 1998; Kraemer \& Crenshaw 2000, hereinafter KC00). 

Although photoionization by the AGN is generally able to reproduce the
UV-optical emission line spectrum of the narrow line region of NGC 1068 (see
e.g. KC00), several of the UV emission lines are substantially at odds
with the predictions of photoionization models that use a MBED of
electron energies. Interestingly, KC00 noted that in the NLR of NGC
1068 the observed CII 1335 flux, relative to H$\beta$ (or CII]
$\lambda$2326), is about an order of magnitude higher than predicted
by the photoionization models that provide a good fit to the overall
UV-optical line spectrum. The presence of this CII
problem in NGC 1068 suggests that it may in fact be a general problem
for AGN photoionized regions, rather than being a problem associated
only with powerful quasars. 

By including the effects of resonant scattering of the AGN's
anisotropic continuum emission, KC00 were able to enhance the relative
flux of the CII $\lambda$1335 line to match the observed
value. Resonant scattering has also been invoked by Ferguson et
al. (1995) in order to explain the larger than expected CIII
$\lambda$977 / CIII] $\lambda$ 1909 and NIII $\lambda$990 / NIII]
$\lambda$1749 flux ratios measured by Kriss et al. (1992) in the NLR
of NGC 1068. 

As discussed in $\S$~\ref{kappa}, our photoionization models using
$\kappa$-distributed electron energies are able to produce a
substantial enhancement in the flux of CII $\lambda$1335 relative to
CII] $\lambda$2326 and H$\beta$. Interestingly, these models also
predict a strong enhancement in  CIII $\lambda$977 / CIII] $\lambda$
1909 and NIII $\lambda$990 / NIII] $\lambda$1749 ratios. Thus, as a
potential new alternative to shock excitation (Kriss et al. 1992) or
resonant scattering (Ferguson et al. 1995; KC00), we propose
$\kappa$-distributed electron energies as the reason for the
discrepancy between observations and photoionization models for these 
CII, CIII and NIII lines. Detailed modeling of the narrow line region
of NGC 1068 using a $\kappa$ distribution is beyond the scope of this
paper, but will be presented in a future work. 

\section{Discussion}
\label{disc}
While investigating the CII $\lambda$1335 / CII] $\lambda$2326
emission line ratio as a potential diagnostic of the temperature,
metallicity and excitation in ionized nebulae at high redshift, we
have identified a failure of canonical photoionization models of 
the narrow-line emitting ionized gas associated with several type 2,
radio-loud quasars at z$\sim$2.5: the models substantially
under-produce CII $\lambda$1335 relative to CII] $\lambda$2326. Given 
that the narrow line region of NGC 1068 shows a similar CII 
problem to the quasars (KC00), we suggest that this may be a general 
problem in fitting the narrow line emitting gases of active galaxies,
rather than being an issue specific only to powerful quasars. 

We have considered several potential causes to explain the higher than
predicted CII $\lambda$1335 / CII] $\lambda$2326 ratio in the type 2
quasars: substantially sub-solar gas metallicity; ionization by
shocks; continuum fluorescence; or $\kappa$-distributed electron
energies. We consider the hypothesis involving $\kappa$-distributed
electron 
energies to be the most promising. It represents what we believe to be
the simplest solution to the CII problem, insofar as it does not
require the presence of a ionizing sources other than the radiation
field of the AGN, nor does it require the summation of multiple models
with different input parameters. Moreover, $\kappa$-distributed
electron energies have been shown to be successful in solving other,
long-standing problems in stellar-photoionized HII regions (see
Nicholls et al. 2012, 2013; Binette et al. 2012), and are already
widely used in studies of plasmas in the solar system (e.g. Livadiotis
\& McComas 2011; Livadiotis et al. 2011). 

We find continuum fluorescence through CII $\lambda$1335, which
enhances the flux of the line above that produced by simple
AGN-photoionization, to be another interesting possibility. Given the
presence of singly ionized carbon 
within the beams of AGN UV continuum radiation fields, it seems
quite likely that CII $\lambda$1335 (and other resonant lines)
could undergo some enhancement by this process. However,  the strength 
of this effect depends on several difficult to constrain properties,
such as the three-dimensional geometry of gas in the host galaxy, and
the UV continuum luminosity of the hidden quasar nucleus. 

The z$\sim$2.5 type 2 quasars considered in this work are all
radio-loud, and are near the top of the radio luminosity function for
AGNs (e.g. Miley \& De Breuck 2008). There is a growing body of
evidence to suggest that at least in some radio-loud active galaxies,
shocks may contribute to the ionization of the extended narrow line
emitting gas (e.g. Clark et al. 1998; Villar-Mart\'{i}n et al. 1999;
Best, R\"ottgering \& Longair 2000; Bicknell et al. 2000; De Breuck et
al. 2000; Inskip et al. 2002; H08; Humphrey et al. 2010). As such,
it would seem entirely reasonable for shocks to contribute to
the ionization of the extended narrow emission line regions of the
distant quasars we have considered herein. However, this
would need to be additional to a strong contribution from
photoionization by the central AGN, in order to explain the flux
ratios of the bright optical emission lines such as [OIII]
$\lambda\lambda$4959,5007, [NeV] $\lambda$3426, H$\beta$, etc (see,
e.g., H08). 

Low gas metallicity is the least appealing of the hypotheses we have
considered. Although the AGN-photoionization models using low gas
metallicity are able to match the observed CII $\lambda$1335 to CII]
$\lambda$2326 flux ratios in the type 2 quasars, which is due to
having higher electron temperature, we point out that low metallicities
are at odds with the findings of several earlier studies which
concluded that the UV-optical emission line spectra of type 2 quasars
are best explained by photoionization models with solar or super-solar
gas metallicities (e.g., Robinson et al. 1987; V01; H08;
Villar-Mart\'{i}n et al. 2008). The presence of pockets of 
low metallicity gas, within higher metallicity gaseous halos
(Tornatore et al. 2007; Cassata et al. 2013) may provide a means to
explain the apparently discrepant implied gas metallicities. 

\section*{Acknowledgments}
AH acknowledges a Marie Curie Fellowship co-funded by the 7$^{th}$
Research Framework Programme and the Portuguese Funda\c{c}\~ao para a
Ci{\^e}ncia e a Tecnologia.  LB acknowledges support from CONACyT
grant CB-128556. We thank the anonymous referee for suggestions that
improved our manuscript.

\section*{Appendix A1: AGN photoionization model grid using a $\kappa$-distribution}
Here we describe the grid of photoionization models with
$\kappa$-distributed electron energies that were computed during the 
preparation of this paper. All models adopt a plane parallel,
single-slab geometry with a constant hydrogen density of 100
cm$^{-3}$, illuminated by an ionizing continuum of spectral shape $S_v
\propto v^{-1.5}$ with a high-energy cut-off of 5$\times$10$^{4}$
eV. Computation was terminated once the hydrogen ionization
fraction fell to 1 per cent of its maximum. Input parameters were
varied as follows; $\kappa$ was set to 2.5, 5, 10, 20, 40, or 80;
ionization parameter U was varied from 0.00001 to 0.20 in steps of
factor 3; the gas metallicity was set to solar, half solar, or
twice solar, with all metal abundances scaled linearly with O/H, except
for N/H which we scaled from its solar value as N/O $\propto$ O/H to
represent secondary nitrogen enrichment. For a subset of our $\kappa$
model grid we show fluxes of various emission lines, normalized to
H$\beta$, in Table A1. For completeness, in Table A2 we also show
output from models with identical input parameters except for 
$\kappa$, which we set to $\infty$ to simulate a Maxwell-Boltzmann
distribution of electron energies. The full grid of models described
in this appendix is available online. 

\begin{table*}
\tiny
\renewcommand\thetable{A1}
\centering
\caption{A sample of our grid of photoionization models using a
  $\kappa$-distribution of electron energies, for $\kappa$ = 40, 20, 10
  and Log U = -3.6, -2.5, -1.7, -0.7. The emission line
  intensities are relative to that of H$\beta$. The full grid is
  available online.} 
\begin{tabular}{lllllllllllll}
\hline
$\kappa$ & 40 & 40 & 40 & 40 & 20 & 20 & 20 & 20 & 10 & 10 & 10 & 10 \\
Log U & -3.6 & -2.6 & -1.7 & -0.7 & -3.6 & -2.6 & -1.7 & -0.7 & -3.6 & -2.6 & -1.7 & -0.7 \\
\hline
C III     977 & 
            3.51e-03
 & 
            4.19e-02
 & 
            3.74e-01
 & 
            2.19e-01
 & 
            1.13e-02
 & 
            1.18e-01
 & 
            6.28e-01
 & 
            3.00e-01
 & 
            1.59e-02
 & 
            4.20e-01
 & 
            1.42e+00
 & 
            5.73e-01
 \\ 
N III     990 & 
            4.39e-05
 & 
            1.77e-03
 & 
            1.49e-02
 & 
            8.77e-03
 & 
            1.61e-04
 & 
            5.00e-03
 & 
            2.50e-02
 & 
            1.12e-02
 & 
            7.00e-05
 & 
            1.78e-02
 & 
            5.66e-02
 & 
            1.84e-02
 \\ 
O VI    1036 & 
            1.61e-09
 & 
            5.31e-05
 & 
            2.18e-01
 & 
            7.85e+00
 & 
            5.37e-09
 & 
            1.23e-04
 & 
            2.97e-01
 & 
            8.71e+00
 & 
            3.92e-11
 & 
            3.47e-04
 & 
            4.65e-01
 & 
            1.02e+01
 \\ 
C II    1037 & 
            2.09e-03
 & 
            2.50e-03
 & 
            1.61e-03
 & 
            3.94e-04
 & 
            4.95e-03
 & 
            5.41e-03
 & 
            2.69e-03
 & 
            6.25e-04
 & 
            1.44e-02
 & 
            1.48e-02
 & 
            5.67e-03
 & 
            1.15e-03
 \\ 
Si III    1207 & 
            1.40e-03
 & 
            8.81e-03
 & 
            4.44e-03
 & 
            8.31e-04
 & 
            3.00e-03
 & 
            1.72e-02
 & 
            6.35e-03
 & 
            1.05e-03
 & 
            1.02e-04
 & 
            3.86e-02
 & 
            1.04e-02
 & 
            1.41e-03
 \\ 
H I    1216 & 
            2.95e+01
 & 
            2.90e+01
 & 
            2.69e+01
 & 
            2.59e+01
 & 
            3.14e+01
 & 
            3.04e+01
 & 
            2.76e+01
 & 
            2.62e+01
 & 
            3.62e+01
 & 
            3.18e+01
 & 
            2.85e+01
 & 
            2.66e+01
 \\ 
N V    1240 & 
            6.32e-07
 & 
            7.58e-04
 & 
            2.12e-01
 & 
            1.02e+00
 & 
            1.43e-06
 & 
            1.39e-03
 & 
            2.62e-01
 & 
            1.09e+00
 & 
            3.33e-08
 & 
            2.93e-03
 & 
            3.60e-01
 & 
            1.22e+00
 \\ 
Si II    1260 & 
            2.03e-02
 & 
            1.81e-02
 & 
            8.44e-03
 & 
            2.47e-03
 & 
            3.42e-02
 & 
            2.64e-02
 & 
            1.06e-02
 & 
            3.04e-03
 & 
            4.17e-02
 & 
            4.34e-02
 & 
            1.41e-02
 & 
            3.74e-03
 \\ 
Si II    1309 & 
            6.78e-03
 & 
            5.89e-03
 & 
            2.70e-03
 & 
            8.03e-04
 & 
            1.08e-02
 & 
            8.15e-03
 & 
            3.25e-03
 & 
            9.43e-04
 & 
            1.14e-02
 & 
            1.25e-02
 & 
            4.05e-03
 & 
            1.09e-03
 \\ 
C II    1335 & 
            6.52e-02
 & 
            6.16e-02
 & 
            2.92e-02
 & 
            8.69e-03
 & 
            9.75e-02
 & 
            8.85e-02
 & 
            3.63e-02
 & 
            1.01e-02
 & 
            1.10e-01
 & 
            1.44e-01
 & 
            5.00e-02
 & 
            1.16e-02
 \\ 
Si IV    1397 & 
            1.83e-03
 & 
            1.03e-01
 & 
            1.19e-01
 & 
            3.69e-02
 & 
            3.24e-03
 & 
            1.74e-01
 & 
            1.76e-01
 & 
            5.26e-02
 & 
            1.82e-04
 & 
            3.25e-01
 & 
            3.13e-01
 & 
            9.13e-02
 \\ 
O IV    1400 & 
            1.37e-05
 & 
            1.01e-02
 & 
            4.29e-01
 & 
            7.55e-01
 & 
            2.59e-05
 & 
            1.59e-02
 & 
            4.99e-01
 & 
            7.75e-01
 & 
            3.70e-07
 & 
            2.76e-02
 & 
            6.23e-01
 & 
            7.77e-01
 \\ 
N IV    1486 & 
            2.07e-05
 & 
            7.17e-03
 & 
            2.26e-01
 & 
            3.37e-01
 & 
            3.59e-05
 & 
            1.08e-02
 & 
            2.65e-01
 & 
            3.74e-01
 & 
            6.86e-07
 & 
            1.77e-02
 & 
            3.41e-01
 & 
            4.53e-01
 \\ 
C IV    1549 & 
            4.18e-04
 & 
            9.74e-02
 & 
            3.48e+00
 & 
            4.22e+00
 & 
            6.63e-04
 & 
            1.40e-01
 & 
            4.10e+00
 & 
            5.04e+00
 & 
            2.32e-05
 & 
            2.11e-01
 & 
            5.29e+00
 & 
            7.02e+00
 \\ 
Ne V    1575 & 
            6.40e-09
 & 
            2.74e-05
 & 
            7.79e-03
 & 
            3.05e-02
 & 
            1.01e-08
 & 
            3.77e-05
 & 
            8.49e-03
 & 
            3.08e-02
 & 
            1.06e-10
 & 
            5.34e-05
 & 
            9.50e-03
 & 
            3.08e-02
 \\ 
Ne IV    1602 & 
            3.59e-06
 & 
            4.29e-04
 & 
            7.60e-03
 & 
            7.75e-03
 & 
            5.52e-06
 & 
            5.96e-04
 & 
            8.50e-03
 & 
            8.06e-03
 & 
            2.01e-07
 & 
            8.67e-04
 & 
            1.00e-02
 & 
            8.42e-03
 \\ 
He II    1640 & 
            7.75e-01
 & 
            1.35e+00
 & 
            1.53e+00
 & 
            1.39e+00
 & 
            7.25e-01
 & 
            1.31e+00
 & 
            1.51e+00
 & 
            1.39e+00
 & 
            2.96e-01
 & 
            1.19e+00
 & 
            1.47e+00
 & 
            1.41e+00
 \\ 
O III    1663 & 
            1.78e-03
 & 
            3.49e-02
 & 
            1.20e-01
 & 
            1.74e-01
 & 
            2.63e-03
 & 
            4.88e-02
 & 
            1.50e-01
 & 
            2.03e-01
 & 
            1.63e-04
 & 
            7.09e-02
 & 
            2.09e-01
 & 
            2.63e-01
 \\ 
N III    1749 & 
            2.92e-03
 & 
            5.61e-02
 & 
            1.42e-01
 & 
            4.03e-02
 & 
            4.04e-03
 & 
            7.50e-02
 & 
            1.67e-01
 & 
            4.60e-02
 & 
            1.82e-04
 & 
            1.02e-01
 & 
            2.14e-01
 & 
            5.72e-02
 \\ 
Mg VI    1806 & 
            7.29e-10
 & 
            5.32e-05
 & 
            4.40e-02
 & 
            7.15e-02
 & 
            9.64e-10
 & 
            6.47e-05
 & 
            4.65e-02
 & 
            7.25e-02
 & 
            0.00e+00
 & 
            7.58e-05
 & 
            4.93e-02
 & 
            7.32e-02
 \\ 
Si II    1809 & 
            6.00e-04
 & 
            4.19e-04
 & 
            1.74e-04
 & 
            5.98e-05
 & 
            6.56e-04
 & 
            4.23e-04
 & 
            1.60e-04
 & 
            5.30e-05
 & 
            2.40e-04
 & 
            4.05e-04
 & 
            1.30e-04
 & 
            3.85e-05
 \\ 
Si III    1890 & 
            4.24e-02
 & 
            1.53e-01
 & 
            4.69e-02
 & 
            8.27e-03
 & 
            4.90e-02
 & 
            1.84e-01
 & 
            4.80e-02
 & 
            7.71e-03
 & 
            4.66e-04
 & 
            2.31e-01
 & 
            4.98e-02
 & 
            6.64e-03
 \\ 
C III    1909 & 
            2.49e-01
 & 
            1.31e+00
 & 
            2.82e+00
 & 
            9.53e-01
 & 
            2.88e-01
 & 
            1.60e+00
 & 
            3.20e+00
 & 
            1.11e+00
 & 
            3.57e-02
 & 
            1.94e+00
 & 
            3.84e+00
 & 
            1.44e+00
 \\ 
N II    2139 & 
            3.87e-02
 & 
            2.74e-02
 & 
            8.23e-03
 & 
            1.67e-03
 & 
            3.93e-02
 & 
            2.70e-02
 & 
            7.32e-03
 & 
            1.35e-03
 & 
            8.22e-03
 & 
            2.43e-02
 & 
            5.73e-03
 & 
            8.81e-04
 \\ 
C II    2326 & 
            5.85e-01
 & 
            3.91e-01
 & 
            1.54e-01
 & 
            6.62e-02
 & 
            5.30e-01
 & 
            3.54e-01
 & 
            1.30e-01
 & 
            5.20e-02
 & 
            1.29e-01
 & 
            2.83e-01
 & 
            9.11e-02
 & 
            2.96e-02
 \\ 
Si II    2335 & 
            3.90e-01
 & 
            2.38e-01
 & 
            9.55e-02
 & 
            3.69e-02
 & 
            3.62e-01
 & 
            2.10e-01
 & 
            7.86e-02
 & 
            2.92e-02
 & 
            7.79e-02
 & 
            1.59e-01
 & 
            5.19e-02
 & 
            1.69e-02
 \\ 
Ne IV    2424 & 
            8.39e-04
 & 
            6.01e-02
 & 
            4.94e-01
 & 
            3.26e-01
 & 
            9.09e-04
 & 
            6.51e-02
 & 
            5.00e-01
 & 
            3.26e-01
 & 
            1.32e-05
 & 
            6.65e-02
 & 
            4.99e-01
 & 
            3.20e-01
 \\ 
O II    2471 & 
            7.58e-02
 & 
            5.20e-02
 & 
            1.42e-02
 & 
            3.02e-03
 & 
            7.31e-02
 & 
            4.98e-02
 & 
            1.28e-02
 & 
            2.57e-03
 & 
            1.05e-02
 & 
            4.20e-02
 & 
            1.02e-02
 & 
            1.88e-03
 \\ 
Mg V    2784 & 
            5.18e-07
 & 
            3.20e-03
 & 
            1.21e-01
 & 
            4.42e-02
 & 
            5.15e-07
 & 
            3.24e-03
 & 
            1.21e-01
 & 
            4.55e-02
 & 
            1.28e-09
 & 
            2.94e-03
 & 
            1.17e-01
 & 
            4.71e-02
 \\ 
Mg II    2800 & 
            1.69e+00
 & 
            1.38e+00
 & 
            8.05e-01
 & 
            3.84e-01
 & 
            1.51e+00
 & 
            1.22e+00
 & 
            6.87e-01
 & 
            3.14e-01
 & 
            4.96e-01
 & 
            8.88e-01
 & 
            4.84e-01
 & 
            1.95e-01
 \\ 
Ne V    3426 & 
            3.59e-06
 & 
            6.27e-03
 & 
            4.69e-01
 & 
            7.62e-01
 & 
            3.39e-06
 & 
            6.07e-03
 & 
            4.50e-01
 & 
            7.44e-01
 & 
            9.28e-09
 & 
            5.15e-03
 & 
            4.04e-01
 & 
            7.00e-01
 \\ 
O II    3727 & 
            6.00e+00
 & 
            3.32e+00
 & 
            8.21e-01
 & 
            1.95e-01
 & 
            5.23e+00
 & 
            2.94e+00
 & 
            6.96e-01
 & 
            1.55e-01
 & 
            5.45e-01
 & 
            2.15e+00
 & 
            4.89e-01
 & 
            9.57e-02
 \\ 
Ne III    3869 & 
            3.89e-01
 & 
            8.75e-01
 & 
            1.14e+00
 & 
            1.08e+00
 & 
            3.41e-01
 & 
            8.27e-01
 & 
            1.13e+00
 & 
            1.09e+00
 & 
            1.92e-02
 & 
            6.78e-01
 & 
            1.05e+00
 & 
            1.10e+00
 \\ 
O III    4363 & 
            5.19e-03
 & 
            7.52e-02
 & 
            1.85e-01
 & 
            2.12e-01
 & 
            5.73e-03
 & 
            8.41e-02
 & 
            2.02e-01
 & 
            2.30e-01
 & 
            1.47e-04
 & 
            8.87e-02
 & 
            2.25e-01
 & 
            2.61e-01
 \\ 
He II    4686 & 
            1.29e-01
 & 
            2.02e-01
 & 
            1.94e-01
 & 
            1.57e-01
 & 
            1.27e-01
 & 
            1.98e-01
 & 
            1.93e-01
 & 
            1.59e-01
 & 
            7.26e-02
 & 
            1.90e-01
 & 
            1.90e-01
 & 
            1.62e-01
 \\ 
O III    5007 & 
            9.92e-01
 & 
            8.76e+00
 & 
            1.40e+01
 & 
            1.33e+01
 & 
            8.82e-01
 & 
            8.23e+00
 & 
            1.35e+01
 & 
            1.32e+01
 & 
            1.39e-02
 & 
            6.66e+00
 & 
            1.23e+01
 & 
            1.27e+01
 \\ 
O I    6300 & 
            7.84e-01
 & 
            4.89e-01
 & 
            2.67e-01
 & 
            1.51e-01
 & 
            5.45e-01
 & 
            3.35e-01
 & 
            1.84e-01
 & 
            1.05e-01
 & 
            4.03e-02
 & 
            1.29e-01
 & 
            7.08e-02
 & 
            3.90e-02
 \\ 
H I    6563 & 
            2.92e+00
 & 
            2.92e+00
 & 
            2.91e+00
 & 
            2.90e+00
 & 
            2.95e+00
 & 
            2.95e+00
 & 
            2.92e+00
 & 
            2.90e+00
 & 
            3.11e+00
 & 
            2.99e+00
 & 
            2.95e+00
 & 
            2.91e+00
 \\ 
N II    6584 & 
            2.38e+00
 & 
            1.06e+00
 & 
            2.77e-01
 & 
            7.57e-02
 & 
            1.99e+00
 & 
            8.98e-01
 & 
            2.23e-01
 & 
            5.61e-02
 & 
            2.38e-01
 & 
            6.01e-01
 & 
            1.36e-01
 & 
            2.73e-02
 \\ 
S II    6724 & 
            1.96e+00
 & 
            8.11e-01
 & 
            5.45e-01
 & 
            3.73e-01
 & 
            1.59e+00
 & 
            6.53e-01
 & 
            4.26e-01
 & 
            2.98e-01
 & 
            3.60e-01
 & 
            3.63e-01
 & 
            2.23e-01
 & 
            1.52e-01
 \\ 
\hline
\end{tabular}
\end{table*}

\begin{table}
\tiny
\renewcommand\thetable{A2}
\centering
\caption{Subset of our model grid using the all the same input
  parameters as those shown in Table A1 except for $\kappa$, which is
  set to infinity in order to simulate a Maxwell-Boltzmann electron
  energy distribution. The emission line
  intensities are relative to that of H$\beta$. The full grid is
  available online.} 
\begin{tabular}{lllll}
\hline
$\kappa$ & $\infty$ & $\infty$ & $\infty$ & $\infty$ \\
Log U & -3.6 & -2.6 & -1.7 & -0.7 \\
\hline
C III     977 & 
            5.95e-04
 & 
            9.49e-03
 & 
            2.02e-01
 & 
            1.61e-01
 \\ 
N III     990 & 
            5.36e-06
 & 
            3.91e-04
 & 
            8.04e-03
 & 
            6.71e-03
 \\ 
O VI    1036 & 
            2.23e-10
 & 
            1.58e-05
 & 
            1.48e-01
 & 
            6.92e+00
 \\ 
C II    1037 & 
            6.09e-04
 & 
            8.96e-04
 & 
            8.49e-04
 & 
            2.13e-04
 \\ 
Si III    1207 & 
            3.98e-04
 & 
            3.40e-03
 & 
            2.76e-03
 & 
            5.82e-04
 \\ 
H I    1216 & 
            2.74e+01
 & 
            2.76e+01
 & 
            2.61e+01
 & 
            2.55e+01
 \\ 
N V    1240 & 
            1.70e-07
 & 
            3.24e-04
 & 
            1.64e-01
 & 
            9.41e-01
 \\ 
Si II    1260 & 
            9.56e-03
 & 
            1.09e-02
 & 
            6.03e-03
 & 
            1.78e-03
 \\ 
Si II    1309 & 
            3.47e-03
 & 
            3.79e-03
 & 
            2.03e-03
 & 
            6.11e-04
 \\ 
C II    1335 & 
            3.61e-02
 & 
            3.87e-02
 & 
            2.18e-02
 & 
            6.70e-03
 \\ 
Si IV    1397 & 
            7.15e-04
 & 
            5.01e-02
 & 
            7.45e-02
 & 
            2.40e-02
 \\ 
O IV    1400 & 
            4.96e-06
 & 
            5.39e-03
 & 
            3.55e-01
 & 
            7.18e-01
 \\ 
N IV    1486 & 
            8.61e-06
 & 
            4.07e-03
 & 
            1.86e-01
 & 
            3.04e-01
 \\ 
C IV    1549 & 
            1.97e-04
 & 
            5.87e-02
 & 
            2.87e+00
 & 
            3.56e+00
 \\ 
Ne V    1575 & 
            3.02e-09
 & 
            1.74e-05
 & 
            6.94e-03
 & 
            2.99e-02
 \\ 
Ne IV    1602 & 
            1.78e-06
 & 
            2.72e-04
 & 
            6.61e-03
 & 
            7.41e-03
 \\ 
He II    1640 & 
            8.03e-01
 & 
            1.38e+00
 & 
            1.54e+00
 & 
            1.39e+00
 \\ 
O III    1663 & 
            9.39e-04
 & 
            2.17e-02
 & 
            9.22e-02
 & 
            1.48e-01
 \\ 
N III    1749 & 
            1.67e-03
 & 
            3.71e-02
 & 
            1.16e-01
 & 
            3.46e-02
 \\ 
Mg VI    1806 & 
            4.36e-10
 & 
            3.95e-05
 & 
            4.08e-02
 & 
            7.04e-02
 \\ 
Si II    1809 & 
            4.95e-04
 & 
            3.96e-04
 & 
            1.81e-04
 & 
            6.33e-05
 \\ 
Si III    1890 & 
            2.97e-02
 & 
            1.16e-01
 & 
            4.36e-02
 & 
            8.43e-03
 \\ 
C III    1909 & 
            1.86e-01
 & 
            9.86e-01
 & 
            2.42e+00
 & 
            7.97e-01
 \\ 
N II    2139 & 
            3.46e-02
 & 
            2.67e-02
 & 
            9.10e-03
 & 
            1.98e-03
 \\ 
C II    2326 & 
            5.98e-01
 & 
            4.19e-01
 & 
            1.77e-01
 & 
            7.90e-02
 \\ 
Si II    2335 & 
            3.89e-01
 & 
            2.60e-01
 & 
            1.11e-01
 & 
            4.40e-02
 \\ 
Ne IV    2424 & 
            6.81e-04
 & 
            5.25e-02
 & 
            4.81e-01
 & 
            3.26e-01
 \\ 
O II    2471 & 
            7.15e-02
 & 
            5.20e-02
 & 
            1.56e-02
 & 
            3.48e-03
 \\ 
Mg V    2784 & 
            4.66e-07
 & 
            3.03e-03
 & 
            1.20e-01
 & 
            4.28e-02
 \\ 
Mg II    2800 & 
            1.74e+00
 & 
            1.48e+00
 & 
            8.98e-01
 & 
            4.41e-01
 \\ 
Ne V    3426 & 
            3.47e-06
 & 
            6.23e-03
 & 
            4.84e-01
 & 
            7.84e-01
 \\ 
O II    3727 & 
            6.42e+00
 & 
            3.63e+00
 & 
            9.50e-01
 & 
            2.39e-01
 \\ 
Ne III    3869 & 
            4.12e-01
 & 
            8.94e-01
 & 
            1.14e+00
 & 
            1.05e+00
 \\ 
O III    4363 & 
            4.09e-03
 & 
            6.26e-02
 & 
            1.65e-01
 & 
            1.93e-01
 \\ 
He II    4686 & 
            1.30e-01
 & 
            2.03e-01
 & 
            1.94e-01
 & 
            1.57e-01
 \\ 
O III    5007 & 
            1.04e+00
 & 
            9.03e+00
 & 
            1.43e+01
 & 
            1.34e+01
 \\ 
O I    6300 & 
            1.02e+00
 & 
            6.51e-01
 & 
            3.52e-01
 & 
            2.00e-01
 \\ 
H I    6563 & 
            2.89e+00
 & 
            2.90e+00
 & 
            2.90e+00
 & 
            2.89e+00
 \\ 
N II    6584 & 
            2.70e+00
 & 
            1.21e+00
 & 
            3.32e-01
 & 
            9.66e-02
 \\ 
S II    6724 & 
            2.26e+00
 & 
            9.37e-01
 & 
            6.40e-01
 & 
            4.42e-01
 \\ 
\hline

\end{tabular}
\end{table}

\end{document}